# Blind ECG Restoration by Operational Cycle-GANs

Serkan Kiranyaz, Ozer Can Devecioglu, Turker Ince, Junaid Malik, Muhammad Chowdhury, Tahir Hamid, Rashid Mazhar, Amith Khandakar, Anas Tahir, Tawsifur Rahman, and Moncef Gabbouj, *Fellow, IEEE*

*Abstract*—**Continuous long-term monitoring of electrocardiography (ECG) signals is crucial for the early detection of cardiac abnormalities such as arrhythmia. Non-clinical ECG recordings acquired by Holter and wearable ECG sensors often suffer from severe artifacts such as baseline wander, signal cuts, motion artifacts, variations on QRS amplitude, noise, and other interferences. Usually, a set of such artifacts occur on the same ECG signal with varying severity and duration, and this makes an accurate diagnosis by machines or medical doctors extremely difficult. Despite numerous studies that have attempted ECG denoising, they naturally fail to restore the actual ECG signal corrupted with such artifacts due to their simple and naive noise model. In this study, we propose a novel approach for blind ECG restoration using cycle-consistent generative adversarial networks (Cycle-GANs) where the quality of the signal can be improved to a clinical level ECG regardless of the type and severity of the artifacts corrupting the signal. To further boost the restoration performance, we propose 1D operational Cycle-GANs with the *generative* neuron model. The proposed approach has been evaluated extensively using one of the largest benchmark ECG datasets from the China Physiological Signal Challenge (CPSC-2020) with more than one million beats. Besides the quantitative and qualitative evaluations, a group of cardiologists performed medical evaluations to validate the quality and usability of the restored ECG, especially for an accurate arrhythmia diagnosis.**

## I. INTRODUCTION

Holter or wearable ECG monitoring has been increasingly used to monitor heart activity for 12 to 48 hours or even longer periods. The extended period of recording time is beneficial for observing sporadic cardiac arrhythmias which would not be possible to diagnose in a shorter time. Doctors recommend patients to avoid sudden movements and high-impact workouts such as running while recording. Even if patients avoid those movements, during their daily routine motion-related slip of the sensor or other interference can induce severe artifacts such as baseline wander, signal cuts, motion artifacts, diminished QRS amplitude, noise, and other interferences. Some typical examples of such corrupted ECG recordings from the benchmark China Physiological Signal Challenge (CPSC-2020) dataset [1] are shown in Figure 1: As can be seen in the figure, the severity of such blended artifacts makes some of the ECG signals undiagnosable by machines or even experienced doctors.

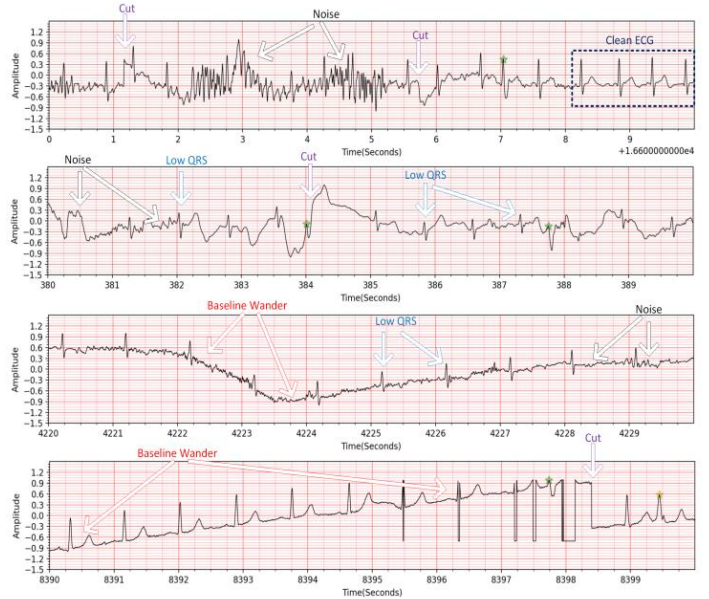

Figure 1: Four 10 second segments from the CPSC-2020 dataset. Arrows with different colors show some typical artifacts.

Even though noise is just one of the artifact types corrupting the ECG signal, numerous studies in the literature address this as the sole denoising problem, and many of which assumed a certain type of (e.g., additive *Gaussian*) noise independent from the signal. To date, several DSP methods from statistical filters or transform-domain denoising [2]-[5] to recent denoising techniques by deep learning have been proposed for ECG denoising. Chieang *et al.* [6] proposed a denoising autoencoder architecture using a fully convolutional network which can be applied to reconstruct the clean data from its noisy version. A 13-layer autoencoder model was applied to MIT-BIH Arrhythmia and Noise Stress datasets corrupted with additive *Gaussian* noise, yielding around 16%, 14%, and 11% SNR (dB) improvements corresponding to the input -1 dB, 3dB, and 7 dB SNR values, respectively. Hamad *et al.* [7] developed a deep learning autoencoder to denoise ECG signals from the discrete wavelet transform coefficients of the ECG signal. The proposed system consists of two stages which are isolating the approximation and thresholding the subband coefficients that will then be used as input to a 14-layer autoencoder to reconstruct a clean signal. They obtained a 6.26 dB SNR improvement on the MIT-BIH Arrhythmia database corrupted with additive *Gaussian* noise. In [8], a deep recurrent neural network (DRNN) model which is a specific hybrid of DRNN and denoising AE is applied to denoising of ECG signal. Both real and synthetic data are used to get improved performance. A new ECG denoising framework based on the generative adversarial network (GAN) is proposed in [9]. For adversarial training of the generative model, both the clean and noisy ECG samples (additive *Gaussian* noise) from the MIT-BIH Arrhythmia database are used. The improved performance of the proposed system over the existing AE-based framework is shown through testing at multiple noise conditions for 5 and 10 dB SNRs.

It is straightforward to develop such supervised ML-based denoising solutions when a clean ECG signal is corrupted by artificial (additive) noise with a fixed type and variance and then turn this as a regression problem by using noisy/clean signal as the input/output of the network, which will eventually learn to suppress the noise. However, such denoising solutions obviously will fail to restore any actual ECG signal from a wearable device corrupted with a blend of artifacts, as typical samples shown in Fig 1. Even only for the "denoising" purpose, assuming an additive and independent noise model with a fixed noise variance is far from being realistic. As can be seen in the ECG segment at the 1st row in Fig 1, the noise level may vary in time and it may neither be additive nor independent from the signal. Therefore, in this study, we address this problem as a *blind* restoration approach thus avoiding any prior assumption over the artifact types and severities. We neither turn it to be a *supervised* regression problem since one cannot have the corrupted and clean ECG signal at the same time in reality unless the artifacts are *artificially* created. That is why, for training, we want to use the *real* corrupted signals with any blend of artifacts, and the network should be able to restore the clean signal while preserving the main characteristics of the ECG patterns. The proposed approach learns to perform transformations between the "clean" (e.g., close to the clinical ECG quality) and the "corrupted" ECG segments using 1D convolutional and operational Cycle-GANs.

Since its first introduction in 2014, GANs [19] and their variations brought a new perspective to the machine learning communities with their superiority in different image synthesis problems. Cycle-Consistent Adversarial Networks (Cycle-GANs) [20] are developed and used for image-to-image translation on unpaired datasets. To accomplish the aforementioned objective, in this study, we first selected batches of clean and corrupted ECG segments from the CPSC-2020 dataset. Then, we adapted the 1D version of Cycle-GANs that can learn to transform the ECG signals (segments) from different batches as the *baseline* method. Since Cycle-GANs can preserve major "patterns" of the signal transformed to the "other" category when a corrupted ECG segment is transformed to a clean segment, the main ECG characteristics (e.g., the interval and timing of R-peaks, QRS waveform of ECG beats, etc.) will still be preserved whilst the quality will be improved. In order to further boost the restoration performance and reduce the complexity, operational Cycle-GANs are proposed in this study. Derived from Generalized Operational Perceptrons [10]-[15], Operational Neural Networks (ONNs) [16]-[18], and their new variants, Self-Organized Operational Neural Networks (Self-ONNs) [21], [22], [30]-[32], are heterogeneous network models with a non-linear neuron model. Self-ONNs are heterogeneous network models with a non-linear neuron model which have shown superior diversity and increased learning capabilities. Recently, Self-ONNs have been shown to outperform their predecessors, CNNs, in many regression and classification tasks. To reflect this superiority in ECG restoration, the convolutional layers/neurons of the native 1D Cycle-GANs are replaced by operational/generative layers/neurons of the Self-ONNs. Once a 1D operational Cycle-GAN is trained over the batches, the generator Self-ONN trained for the "corrupted" to "clean" ECG segment transformation can then be used for the ECG restoration. The performance is evaluated over the SCPC-2020 dataset quantitatively by the performance comparisons using the benchmark peak detectors, Pan and Tompkins [23] and Hamilton [24], qualitatively (visually), and also by the medical doctors for arrhythmia diagnosis.

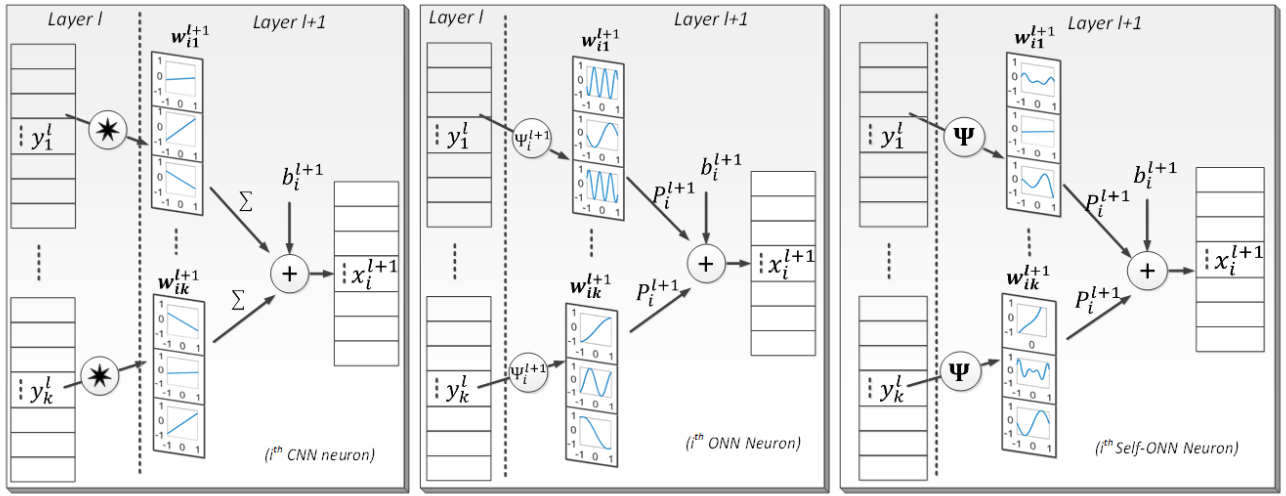

Figure 2: Depiction of the 1D nodal operations with the 1D kernels of the $i$th neuron of CNN (left), ONN (middle), and Self-ONN (right).

The rest of the paper is organized as follows: a brief outline of 1D Self-ONNs and the proposed approach with the operational Cycle-GANs are introduced in Section II. The results are presented in Section III. Finally, Section IV concludes the paper and suggests topics for future research.

## II. PROPOSED APPROACH

In this section, we first briefly summarize Self-ONNs and their main properties. Then we introduce the proposed approach by 1-D Self Operational Cycle GANs for ECG restoration.

### A. 1D Self-Organized Operational Neural Networks

In this section, we introduce the main network characteristics of 1D Self-ONNs[1] with the formulation of forward propagation. Figure 2 shows 1D nodal operations of a CNN, ONN with fixed (static) nodal operators, and Self-ONN with *generative* neuron which can have any arbitrary nodal function, **Ψ**, (including possibly standard types such as linear and harmonic functions) for each kernel element of each connection. Obviously, Self-ONN has the potential to achieve greater operational diversity and flexibility, allowing any nodal operator function to be formed without the use of an operator set library or a prior search process to select the best nodal operator.

The kernel elements of each generative neuron of a Self-ONN perform any nonlinear transformation, $\psi$, the function of which can be expressed by the Taylor-series near the origin ($a = 0$),

$$\psi(x) = \sum_{n=0}^{\infty} \frac{\psi^{(n)}(0)}{n!} x^n \qquad (1)$$

The $Q^{th}$ order truncated approximation, formally known as the Taylor polynomial, takes the form of the following finite summation:

$$\psi(x)^{(Q)} = \sum_{n=0}^{Q} \frac{\psi^{(n)}(0)}{n!} x^n \qquad (2)$$

The above formulation can approximate any function $\psi(x)$ near 0. When the activation function bounds the neuron's input feature maps in the vicinity of 0 (e.g., *tanh*), the formulation in (2) can be exploited to form a composite nodal operator where the power coefficients, $\frac{\psi^{(n)}(0)}{n!}$, can be the parameters of the network learned during training.

It was shown in [21], [22], and [29] that the nodal operator of the $k^{th}$ generative neuron in the $l^{th}$ layer can take the following general form:

$$\widetilde{\psi_k^l}\left(w_{ik}^{l(Q)}(r), y_i^{l-1}(m+r)\right)$$
$$= \sum_{q=1}^{Q} w_{ik}^{l(Q)}(r,q)\left(y_i^{l-1}(m+r)\right)^q \qquad (3)$$

Let $x_{ik}^l \in \mathbb{R}^M$ be the contribution of the $i^{th}$ neuron's at the $(l-1)^{th}$ layer to the input map of the $l^{th}$ layer. Therefore, it can be expressed as,

$$\widetilde{x_{ik}^l}(m) = \sum_{r=0}^{K-1} \sum_{q=1}^{Q} w_{ik}^{l(Q)}(r,q)\left(y_i^{l-1}(m+r)\right)^q \qquad (4)$$

where $y_i^{l-1} \in \mathbb{R}^M$ is the output map of the $i^{th}$ neuron's at the $(l-1)^{th}$ layer, $w_{ik}^{l(Q)}$ is a learnable kernel of the network, which is a $K \times Q$ matrix, i.e., $w_{ik}^{l(Q)} \in \mathbb{R}^{K \times Q}$, formed as, $w_{ik}^{l(Q)}(r) = [w_{ik}^{l(Q)}(r,1), w_{ik}^{l(Q)}(r,2), \dots, w_{ik}^{l(Q)}(Q)]$. By the commutativity of the summation operations in (4), one can alternatively write:

$$\widetilde{x_{ik}^l}(m) = \sum_{q=1}^{Q} \sum_{r=0}^{K-1} w_{ik}^{l(Q)}(r, q-1) y_i^{l-1}(m+r)^q \qquad (5)$$

One can simplify this as follows:

$$\widetilde{x_{ik}^l} = \sum_{q=1}^{Q} Conv1D\left(w_{ik}^{l(Q)}, \left(y_i^{l-1}\right)^q\right) \qquad (6)$$

Hence, the formulation can be accomplished by applying Q 1D convolution operations. Finally, the output of this neuron can be formulated as follows:

$$x_k^l = b_k^l + \sum_{i=0}^{N_{l-1}} x_{ik}^l \qquad (7)$$

where $b_k^l$ is the bias associated with this neuron. The $0^{th}$ order term, $q = 0$, the DC bias, is ignored as its additive effect can be compensated by the learnable bias parameter of the neuron. With the $Q = 1$ setting, a *generative* neuron reduces back to a convolutional neuron.

The raw-vectorized formulations of the forward propagation, and detailed formulations of the Back-Propagation (BP) training in raw-vectorized form can be found in [22] and [29].

### B. 1D Operational Cycle-GANs

The general framework of our proposed ECG restoration scheme is shown in Figure 3. We follow a segment-based restoration scheme where each ECG segment has 10 seconds duration. With the sampling frequency of 400Hz, this corresponds to $m = 4000$ samples per segment. By visual evaluation, we have carefully selected the batches of 4000 clean and corrupted ECG segments to establish the training dataset. A segment is "clean" only when there is no visible sign of *any* artifact; otherwise, it is a "corrupted" segment. CPCS-2020 dataset has supraventricular ectopy (S) and ventricular ectopy (V) type beats. In order to ensure an unbiased training with respect to the presence of any arrhythmia, for both clean and corrupted segments, the selection is performed in such a way that 33.4% of them contain at least one arrhythmia type. Moreover, to ensure an unbiased training on the type and severity of the corruption, such corrupted segments with different (blend of) artifacts (e.g., different noise types/levels, baseline wander, cuts, QRS amplitude shrinkage, etc.) and with different severity levels are selected.

---

[1] The optimized PyTorch implementation of Self-ONNs is publicly shared in http://selfonn.net/ .

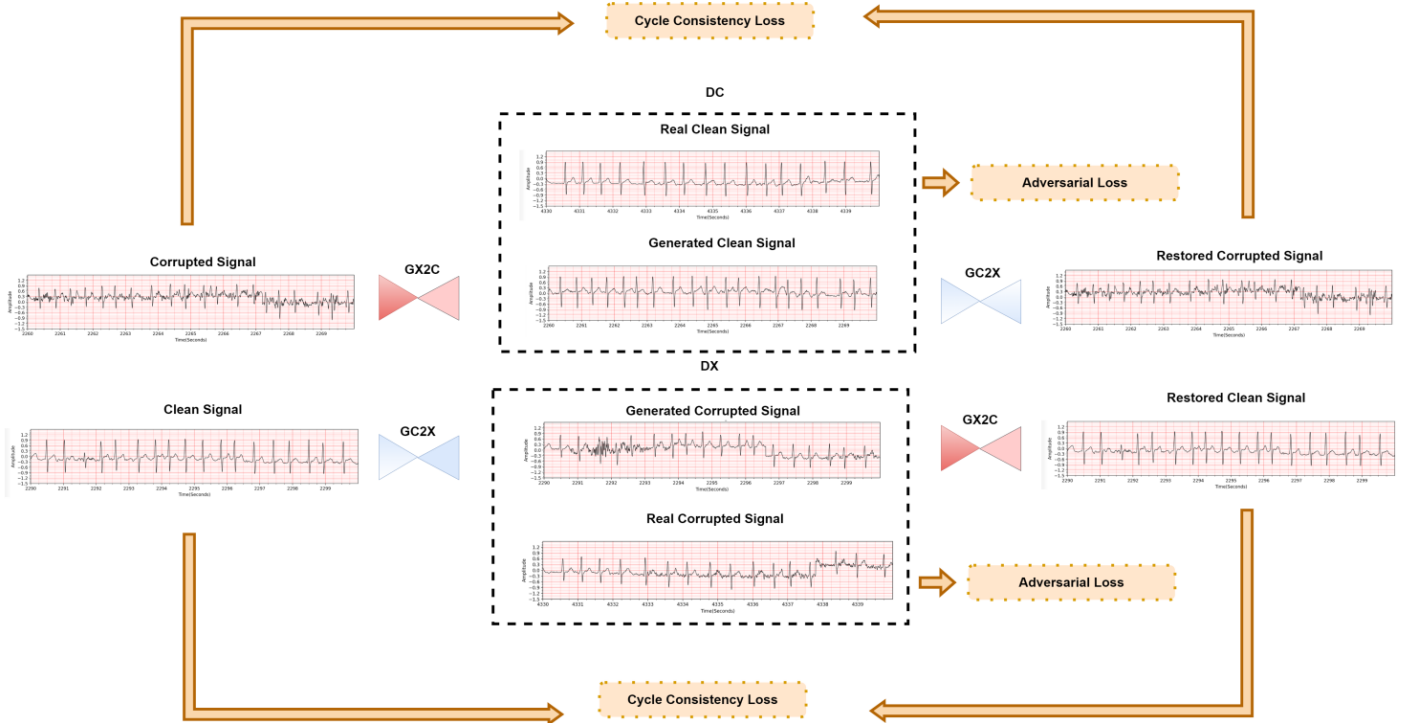

Figure 3: The proposed ECG restoration approach using operational Cycle-GANs.

Once the training dataset is formed, we adapted the 1D version of Cycle-GANs that can learn to transform the ECG signals (segments) from different batches as the *baseline* method. As discussed earlier, Cycle-GANs can preserve the major characteristics of the signal when it is transformed to the "other" category. Therefore, one of the generators will learn to transform the corrupted ECG segments to their "clean" version whilst preserving the main ECG characteristics (e.g., the interval and timing of R-peaks, QRS waveform of an ECG beat, etc.). The most critical point is that the transformation of the arrhythmic beats should be unaltered (temporally or morphologically) besides the quality improvement. In other words, an arrhythmic beat in a corrupted segment should *not* be transformed to a normal beat. This is why the unbiased selection scheme for forming the training set is crucial. This is one of the critical evaluation criteria that will be assessed by a group of cardiologists.

As a new-generation ANN model, Self-ONNs outperform conventional (deep) CNNs on many ML and CV tasks. To reflect this superiority in ECG restoration, the proposed approach for ECG restoration is to use Operational 1D Cycle-GANs where the convolutional layers/neurons of native 1D Cycle-GANs (both the generator and discriminator) are replaced by the operational layers with generative neurons of the Self-ONNs. To reduce the complexity, Operational GANs have four times fewer neurons and around 5 times fewer network parameters than the baseline model. This will also allow us to perform comparative evaluations between CNNs and ONNs in the GAN domain for the first time. As shown in Fig 2, an ECG segment from each batch is randomly selected as the input pair for the Cycle-GAN. They are first linearly normalized into the range of [-1 1], as follows:

$$X_N(i) = \frac{2(X(i) - X_{\min})}{X_{max} - X_{min}} - 1 \tag{8}$$

where $X(i)$ is the original sample amplitude in the segment, $X_N(i)$ is the normalized segment, $X_{min}$ and $X_{max}$ are the minimum and maximum amplitudes within the segment, respectively. The proposed approach consists of two Self-ONN based models: Generator and Discriminator. As in [20], the proposed approach consists of two sets of generators and discriminators. While the generator "corrupted-to-clean" (GX2C) learns to transform a corrupted segment to a clean one, the aim of the generator "clean-to-corrupted" (GC2X) will learn the opposite and will be discarded after the training. Both corresponding discriminators, "corrupted" (DX) and clean (DC) aim to maximize the adversarial loss functions so as to generate more realistic transformations. The loss functions are expressed in Eqs. (9) and (10).

$$Loss_{adv1}(\text{GX2C}, DC, X_X)$$
$$= \frac{1}{m}\sum_{i=1}^{m}(1 - DC(GX2C(X_X(i))))^2 \tag{9}$$

$$Loss_{adv2}(GC2X, DX, X_C)$$
$$= \frac{1}{m}\sum_{i=1}^{m}(1 - DX(GC2X(X_C(i))))^2 \tag{10}$$

where $X_X$ and $X_C$ are the corresponding corrupted and clean ECG segments, respectively. In order to improve the preservation of the ECG characteristics, unlike the traditional GANs, we further use the cycle-consistency loss as expressed in (11).

$$Loss_{cyc}(GX2C, GC2X, X_X, X_C)$$
$$= \frac{1}{m} \sum_{i=1}^{m} \left[ GC2X\left(GX2C(X_X(i))\right) - X_X(i) \right]$$
$$+ \frac{1}{m} \sum_{i=1}^{m} \left[ GX2C\left(GC2X(CS(i))\right) - X_C(i) \right] \quad (11)$$

In addition to adversarial and cycle consistency losses, the identity loss as given in (12) is defined for reducing the level of variation if the class of the input sample is the same as the desired output.

$$Loss_{ide}(GX2C, GC2X, X_X, X_C)$$
$$= \frac{1}{m} \sum_{i=1}^{m} \left[ \left( GX2C(X_C(i)) \right) - X_C(i) \right]$$
$$+ \frac{1}{m} \sum_{i=1}^{m} \left[ \left( GC2X(X_X(i)) \right) - X_X(i) \right] \quad (12)$$

The objective of any Cycle-GAN training is to minimize the total loss in (13).

$$Loss_{total} = Loss_{adv1} + Loss_{adv2} + \lambda \, Loss_{cyc}$$
$$+ \; \beta \, Loss_{ide} \quad (13)$$

The experimental setup and network parameters will be presented in the next section.

## III. EXPERIMENTAL RESULTS

In this section, the benchmark CPSC-2020 dataset will first be introduced. Then, the experimental setup used for the evaluation of the proposed ECG restoration approach will be presented. The comparative evaluations and the overall results of the experiments obtained over real Holter recordings will be presented in the following step. The quantitative, qualitative, and medical evaluations (by a group of cardiologists) are all performed. Additionally, the computational complexity of the proposed approach will be evaluated in detail.

### A. CPSC-2020 Dataset

The China Physiological Signal Challenge 2020, (CPSC-2020) dataset is not only one of the largest benchmark datasets with more than 1M beats, but it also presents natural Holter ECG recordings with actual artifacts discussed earlier and thus, it is ideal for evaluating the proposed approach. The dataset consists of 10 single-lead ECG recordings of 10 arrhythmia patients each of which has a duration of around 24 hours. The details of the dataset are presented in TABLE 1.

TABLE 1 DATASET DETAILS

| PAT. NO | AF PATIENT? | DURATION (HOUR) | NO. OF BEATS | NO. OF V BEATS | NO. OF S BEATS |
|---------|-------------|-----------------|--------------|----------------|----------------|
| A01 | No | 25.89 | 109,062 | 0 | 24 |
| A02 | Yes | 22.83 | 98,936 | 4,554 | 0 |
| A03 | Yes | 24.7 | 137,249 | 382 | 0 |
| A04 | No | 24.51 | 77,812 | 19,024 | 3,466 |
| A05 | No | 23.57 | 94,614 | 1 | 25 |
| A06 | No | 24.59 | 77,621 | 0 | 6 |
| A07 | No | 23.11 | 73,325 | 15,150 | 3,481 |
| A08 | Yes | 25.46 | 115,518 | 2,793 | 0 |
| A09 | No | 25.84 | 88,229 | 2 | 1,462 |
| A10 | No | 23.64 | 72,821 | 169 | 9,071 |

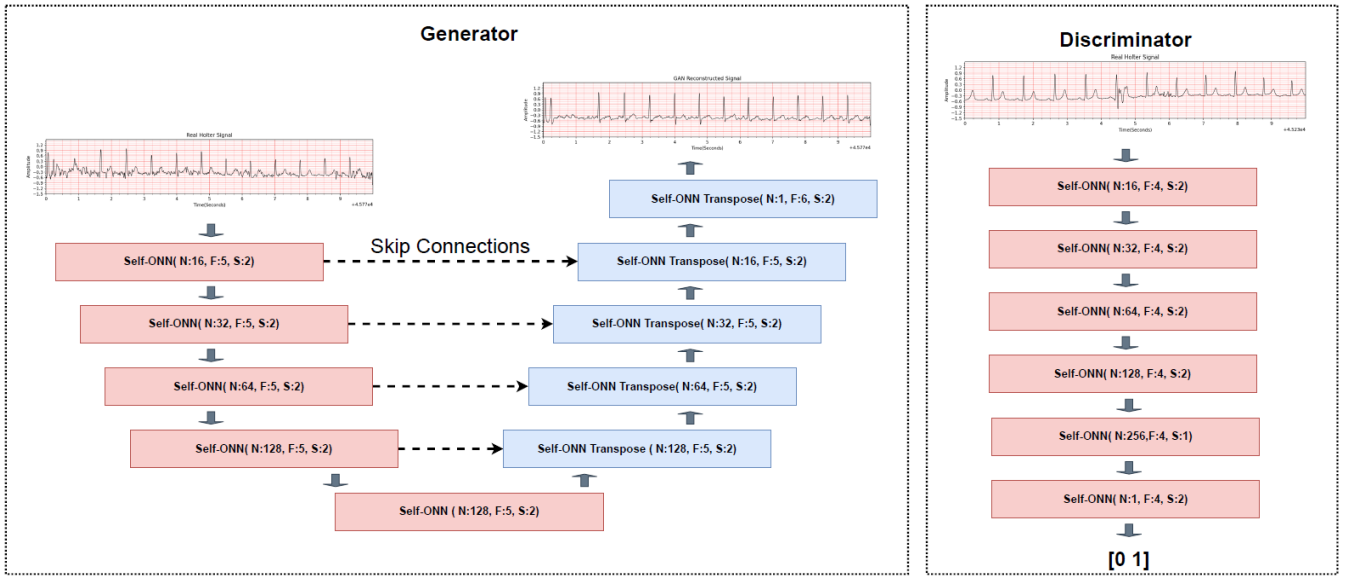

Figure 4: The Generator and Discriminator architectures of the proposed approach.

### B. Experimental Setup

For both generators GX2C and GC2X of both baseline (convolutional) and operational Cycle-GANs, 10-layer U-Net configuration is used with 5 1-D convolutional/operational layers and 5 transposed convolutional/operational with residual connections. The kernel sizes are set as 5 except that the last transposed layer the kernel size 6 is used. The stride is set as 2 for both convolutional/operational and transposed convolutional/operational layers. Both discriminators consist of 6 operational layers with a kernel size of 4. The stride for layers is set as 2, 2, 2, 2, 1, and 2 respectively. As a loss function in the Discriminator, MSE is computed between the discriminator output and label vectors bot of which have dimension 30. The architectures for the generators and discriminators are shown in Figure 4. For all experiments, we employ a training scheme with a maximum of 1000 BP iterations with batch size 8. The Adam optimizer with the learning rate $10^{-5}$ is used for both generators and discriminators. The loss weights $\lambda$ and $\beta$ in (9) are set as 10 and 5. We implemented the proposed 1D Self ONN architectures using the FastONN library [25] based on Python [26] and PyTorch [27]. For the training dataset, 4000 clean and corrupted segments with a duration of 10 seconds (4000 samples) are selected, and the segments from the rest of the data are used for testing and evaluation.

### C. Quantitative Evaluations over Peak Detection

For quantitative evaluation, we use the landmark Hamilton [23], and Pan and Tompkins [24] peak detectors and evaluated the performance gain achieved by the proposed ECG restoration approach. Commonly used performance metrics Precision (Pre), Sensitivity (Sen), F1-Score (F1), and the number of missed S and V arrhythmia beats are used to compare performance. The calculation of True Positives (TP), False Negatives (FN), and False Positives (FP) were taken within a tolerance of ±75 msec [28] of the truth peak location. Since this is an R-peak detection operation, True Negatives (TN) do not exist as a performance measure. The formulations for these performance metrics can be expressed as follows:

$$Pre = \frac{TP}{TP + FP}, \quad Sen = \frac{TP}{TP + FN},$$
$$F1 = \frac{2 Pre Sen}{Pre + Sen} \tag{14}$$

Table 2 and Table 3 present peak detection performances of the landmark detectors over the original and restored ECG segments by the *baseline* and operational Cycle-GANs. Besides the *baseline* model, we also used a more complex Cycle-GAN (*Cycle-GANx4*) with four times more neurons and around 5 times more network parameters than operational Cycle-GANs to evaluate the gain achieved at the expense of higher complexity. Finally, we present the peak detection results over the two-pass restoration by the Self-ONN generator (the output of GX2C is again restored by GX2C a second time).

Both peak detection results clearly show that the peak detection errors (FP and FN) are both reduced over the restored ECG by the proposed approach without exception. When the same network configuration is used (with the same number of neurons), operational Cycle-GANs significantly outperform the *baseline* Cycle-GANs in all metrics. As expected, only when the number of neurons is increased by four times, the complex Cycle-GANx4 can achieve a slightly better performance in overall peak detection (around 0.3 – 0.4 % difference in F1); however, operational Cycle-GANs can still outperform the Cycle-GANx4 in the peak detection of the arrhythmia beats. In fact, the detection of the arrhythmia beats is the most important objective since peak detectors are commonly used as a pre-processing step for the arrhythmia diagnosis by both medical doctors and machines. After the restoration by the operational Cycle-GANs, the number of missing V beats can be reduced by more than 40%. With an additional restoration pass by the operational Cycle-GANs (two-pass), this can be improved by more than 50% for the Pan and Tompkins peak detector.

#### Table 2: Peak Detection Performance Of Hamilton Peak Detector [24].

| | TP | FN | FP | Recall | Precision | F1 | S Missed | V missed |
|---|---|---|---|---|---|---|---|---|
| **Original Signal** | 990647 | 35448 | 62985 | 96.52 | 93.68 | 94.97 | **393** | 6573 |
| **CycleGAN (*baseline*)** | 986712 | 39383 | 46072 | 96.22 | 95.24 | 95.66 | 440 | 6332 |
| ***CycleGANx4*** | 992829 | 33266 | 35016 | 96.77 | 96.30 | 96.50 | 450 | 4370 |
| **Operational CycleGAN(Q=3)** | 992781 | 33314 | 42749 | 96.79 | 95.57 | 96.13 | 401 | 3407 |
| **Operational Cycle-GAN(Q=3) (two-pass)** | 989865 | 36230 | 39880 | 96.53 | 95.84 | 96.13 | 445 | **2981** |

#### Table 3: Peak Detection Performance Of Pan&Tompkins Peak Detector [23].

| | TP | FN | FP | Recall | Precision | F1 | S Missed | V missed |
|---|---|---|---|---|---|---|---|---|
| **Original Signal** | 995494 | 30601 | 29458 | 97.04 | 97.11 | 97.05 | **317** | 4922 |
| **CycleGAN (*baseline*)** | 990395 | 35700 | 21613 | 96.58 | 97.87 | 97.22 | 363 | 4308 |
| ***CycleGANx4*** | 997670 | 28425 | 15617 | 97.26 | 98.47 | 97.86 | 363 | 3323 |
| **Operational CycleGAN(Q=3)** | 995622 | 30473 | 19926 | 97.07 | 98.05 | 97.55 | 344 | **2809** |
| **Operational Cycle-GAN(Q=3) (two-pass)** | 992313 | 33782 | 17859 | 96.77 | 98.23 | 97.49 | 401 | 2864 |

## D. Medical Evaluation

There are two objectives of the medical evaluation:

- To find out the best ECG signal for arrhythmia diagnosis with respect to the cardiologist's perspective.
- To find out whether ECG restoration causes loss of any arrhythmia beats or the creation of false arrhythmia beats.

The first objective is not only to evaluate original vs. restored ECG for arrhythmia diagnosis, but it also serves the purpose to determine which restoration approach will be the most preferable by the cardiologists. The 2nd objective is especially critical for arrhythmia diagnosis since arrhythmia beats are usually rare and hence, they should not be removed, or no false arrhythmia beats should be created by the restoration method.

To accomplish these evaluation objectives, we randomly selected 2000 ECG segments from the test partition of the CPSC-2020 dataset and restored them using the four Cycle-GAN methodologies (baseline, Cycle-GANx4, Operational and Operational two-pass). A group of cardiologists evaluated their outputs and compared them along with the original signal. Their responses are collected in a survey, and we got the following medical evaluation results: Among all doctors' responses, the original and restored ECG signals are found as the best option for arrhythmia detection with 4.49 % and 95.51% of the time, respectively. This clearly shows that ECG restoration is indeed crucial for a better medical evaluation by the doctors. Moreover, they have found only 0.04% of the time where an arrhythmia beat is restored as a normal beat and hence missed. No S beat was missed, and no false arrhythmia beat has ever been created by any of the restoration approaches. This fulfills the 2nd and the most critical objective.

Among the three ECG restoration approaches, Cycle-GANx4, operational Cycle-GAN, and operational Cycle-GAN with two-pass, the doctors have found them the best for diagnosis 28.3 %, 6.9 %, and 64.8 % of the time, respectively. The most favored method is, therefore, operational Cycle-GAN with two-pass, as expected. This outcome is mainly due to the superior restoration quality achieved especially on the arrhythmia beats and the noise suppression level.

## E. Qualitative Evaluation

For the qualitative (visual) evaluation, Figure 5 and Figure 6 show two original ECG segments from the records of patients 2 and 7 in the CPSC-2020 dataset along with the three restored ECG segments by the Cycle-GANs (baseline Cycle-GAN output is omitted since other networks almost always outperform it). 10 more visual results are shown in the Appendix. In the figures, the beats annotated with green and yellow stars correspond to V and S type arrhythmia beats, respectively.

The first and the foremost observation is that the quality of the restored ECG segments has significantly been improved compared to the original ECG segment regardless of the Cycle-GAN type, e.g., the noise has been suppressed significantly or cleaned entirely, the baseline wander or fluctuations are removed, the QRS beat amplitudes are mostly enhanced, the abrupt signal cuts are removed, etc. The proposed restoration approaches succeed to create authentic QRS beats with the right timing with respect to their original counterparts. On the other hand, when the original ECG signal is sufficiently clean, it is kept intact after the restoration without any artificial variations or degradations. Moreover, the arrhythmia beats in the original ECG segment are restored as to the arrhythmia beats with the corresponding type. As discussed earlier, this is critical for arrhythmia diagnosis by both machines and cardiologists.

A closer look at the figures reveals the fact that the best restoration has been performed by the operational Cycle-GANs with two-pass (on the bottom), i.e., the best noise suppression, QRS amplitude restoration, and the removal of baseline wander and cuts. This is in accordance with the medical evaluations by the cardiologists. An interesting observation worth mentioning in Figure 5 is that a possible V-beat was missed by the Chinese cardiologists due to the excess noise; however, after the restoration, it becomes quite straightforward to diagnose this arrhythmia (as shown in the figure with a green arrow). The opposite is also true; due to severe artifacts, the Chinese cardiologists mislabeled a V beat (marked with a green star) as shown in Figure 6 by the red arrow. Only after the restoration, the cardiologists in this study confirm that it should not be a V beat or any beat at all. We present 10 more sample ECG segments and their restoration results in the Appendix (see: Figure 7 - Figure 16). Although the Cycle-GANx4 has a significantly higher number of learning units and complexity, operational Cycle-GANs usually outperform them especially on QRS amplitude restoration (e.g., some R-peaks could not be restored fully by the Cycle-GANx4 as shown in Figure 7 and Figure 8 with blue arrows). Similarly, in Figure 9, the cut is not restored by the Cycle-GANx4 when compared with the restorations by Operational GANs (shown by a purple arrow). Finally, although quite rare, some restoration issues are shown in Figure 17 and Figure 18. In Figure 17, all GAN restorations fail for the two V-beats due to their very low amplitude. In fact, the cardiologists in this study also raise a concern about their validity. In Figure 18, the arrow on the left shows that all GANs over-correct a cut to be restored as a somewhat distorted ECG beat. The arrow on the right, however, shows a cut that probably coincides with an ECG beat (based on the timing). The Cycle-GANx4 removed it completely during restoration while the operational Cycle-GANs restored an ECG beat instead. Once again, it is hard to decide whether it is indeed an ECG beat or not, and hence, this may be an over-correction or a valid restoration.

## F. Computational Complexity

For computational complexity analysis, the network size, total number of parameters (PARs), and inference time (to restore an ECG segment) for each network configuration are computed and reported in Table 4. The detailed formulations of the PARs calculations for Self-ONNs can be found in [25]. All the experiments were carried out on a 2.2 GHz Intel Core i7 with 16 GB of RAM and NVIDIA GeForce RTX 3080 graphic card. For the implementation of the Cycle-GANs and operational Cycle-GANs, Python with PyTorch library is used. Both the training and testing phases of the classifier were processed using GPU cores. As the inference time and PARs indicate, the operational Cycle-GAN is significantly faster and less complex than the Cycle-GANx4.

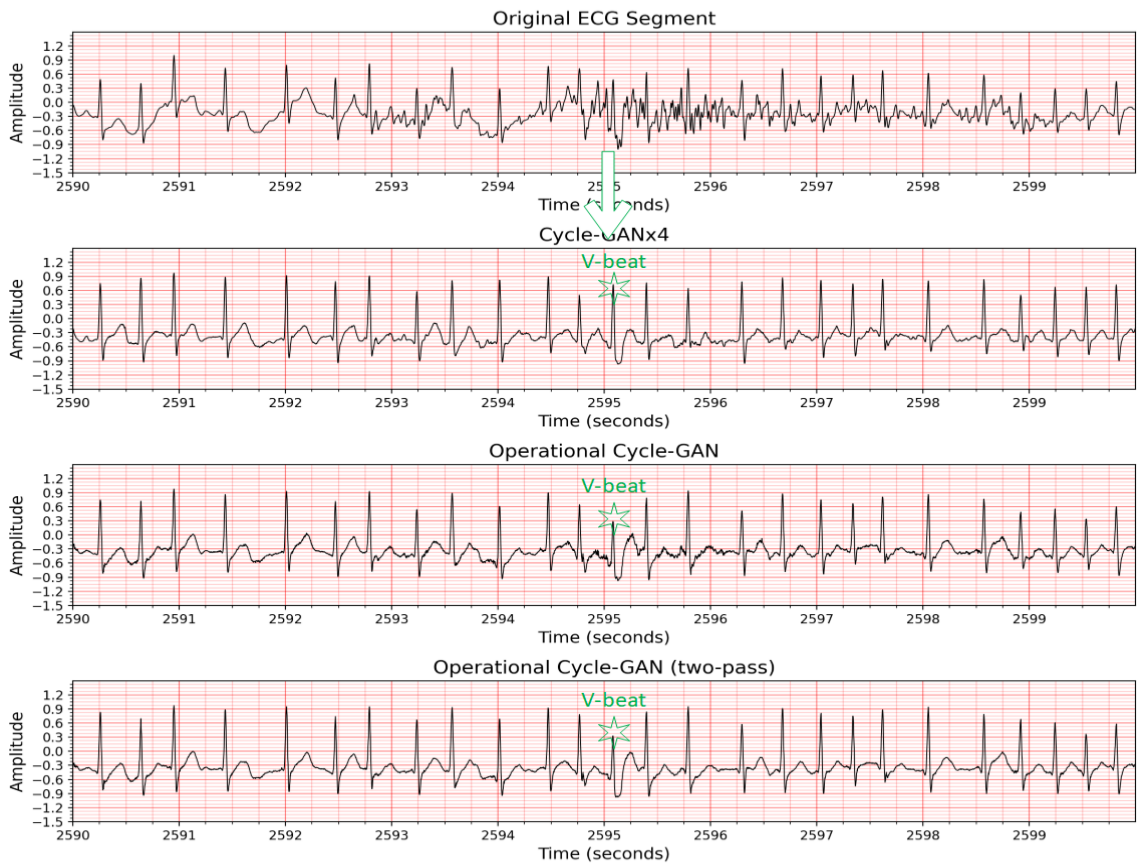

Figure 5: Sample ECG segment from Patient 2 and its corresponding GAN output signals.

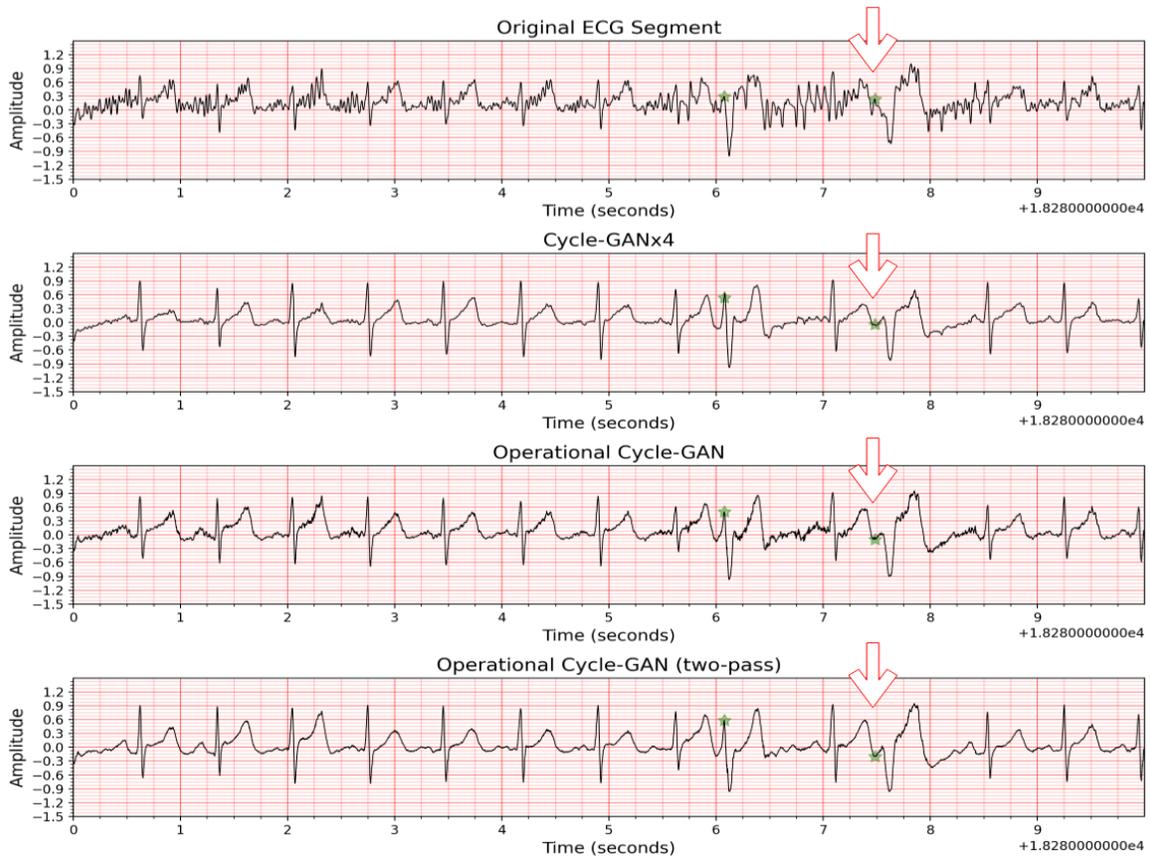

Figure 6: Sample ECG segment from Patient 7 and its corresponding GAN output signals.

TABLE 4 COMPUTATIONAL COMPLEXITY OF THE NETWORKS.

| | PARs (M) | | | Inf. Time (msec) |
|---|---|---|---|---|
| | GX2C | DX | Total | |
| *Baseline* | *0.260* | *0.175* | *0.873* | *0.14* |
| *Cycle-GANx4* | *4.2* | *2.8* | *14* | *3.5* |
| *Operational Cycle-GAN* | *0.781* | *0.544* | *2.7* | *0.32* |
| *Operational Cycle-GAN (two-pass)* | *0.781* | *0.544* | *2.7* | *0.64* |

## IV. CONCLUSION

The major problem of Holter and wearable ECG sensors is that the acquired ECG signal may severely be corrupted by a blend of artifacts, and this makes it too difficult, if not infeasible, to diagnose any heart abnormality by machines or humans. In this study, we propose a novel approach to restore the ECG signal to a clinical level quality regardless of the type or severity of the artifacts. Therefore, we follow a different path from the prior works, which approached this as a "denoising" problem for additive (artificial) noise with a fixed type and power so that they could propose a *supervised* solution. Such common regression-based solutions are not useful in practice and that is why this study addressed this problem with a *blind* restoration approach without any prior assumption over the artifact types and severity. As the baseline method, we proposed 1D Cycle-GANs, and to further boost the performance, we proposed operational Cycle-GANs. Once Cycle-GANs are trained over the clean and corrupted batches, the generator, *GX2C,* learns to transform the corrupted ECG segments to clean counterparts while preserving the ECG characteristics. The optimized PyTorch code and the labeled *CPSC-2020* dataset are publicly shared in [33].

The quantitative, qualitative, and medical evaluations performed over an extensive set of real Holter recordings demonstrate that the corrupted ECG can indeed be restored with a desired (clinical) quality level, which in turn improves the efficiency and accuracy of ECG diagnosis by machines and humans. In particular, the R-peak detection performances of the two landmark detectors have been significantly improved over the restored signal. During the medical evaluation, the cardiologists confirmed that the restored ECG signal is more useful for arrhythmia diagnosis 95.51% of the time. They further note that the restoration has almost no side effects on the arrhythmia beats, i.e., neither causing an arrhythmic beat to turn to a normal beat nor transforming a normal beat into an arrhythmic beat. Finally, besides the superior ECG quality achieved by the proposed restoration approach, the visual evaluation further demonstrated that the hidden/undetected arrhythmia events can possibly be diagnosed from the restored ECG. A similar conclusion can also be made on the significant peak detection performance gain of arrhythmia beats achieved after the restoration. Among all proposed restoration approaches by 1D Cycle-GANs, the novel operational Cycle-GANs have a superior restoration performance and can even outperform a more complex counterpart with convolutional neurons. This is not surprising considering the superiority of Self-ONNs in many challenging ML and CV tasks over the (deep) CNN models [29]-[31].

Despite the elegant restoration performance, we note that very occasionally some potential arrhythmia beats with very low amplitudes may not be distinguished from the background noise, and hence not restored. Moreover, few over-corrections were encountered yielding artificial beats. Such minority cases can be addressed by designing a cost function that incorporates the class information (normal, S, and V type beats). Finally, the depth and complexity of the operational Cycle-GANs can further be reduced while boosting the restoration performance by using the super neuron model recently proposed in [32]. These will be the topics of our future research.

APPENDIX

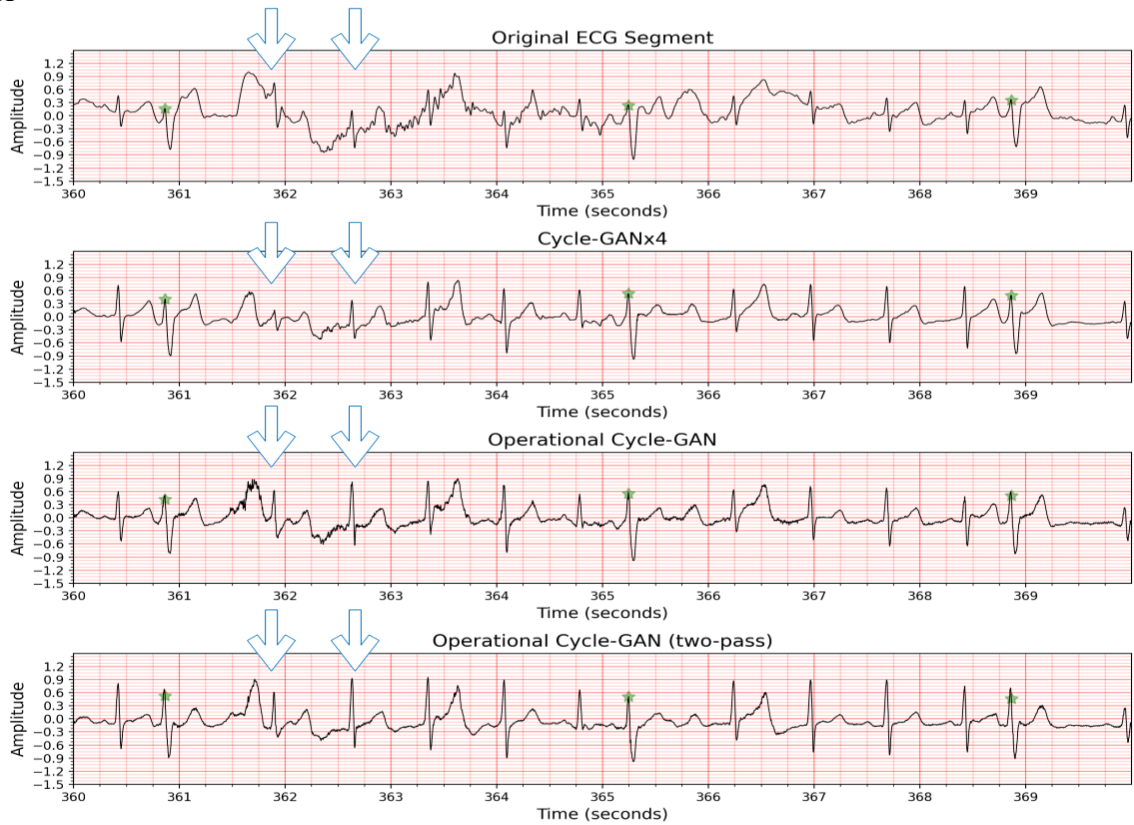

Figure 7: Sample ECG segment from Patient 7 and its corresponding GAN output signals.

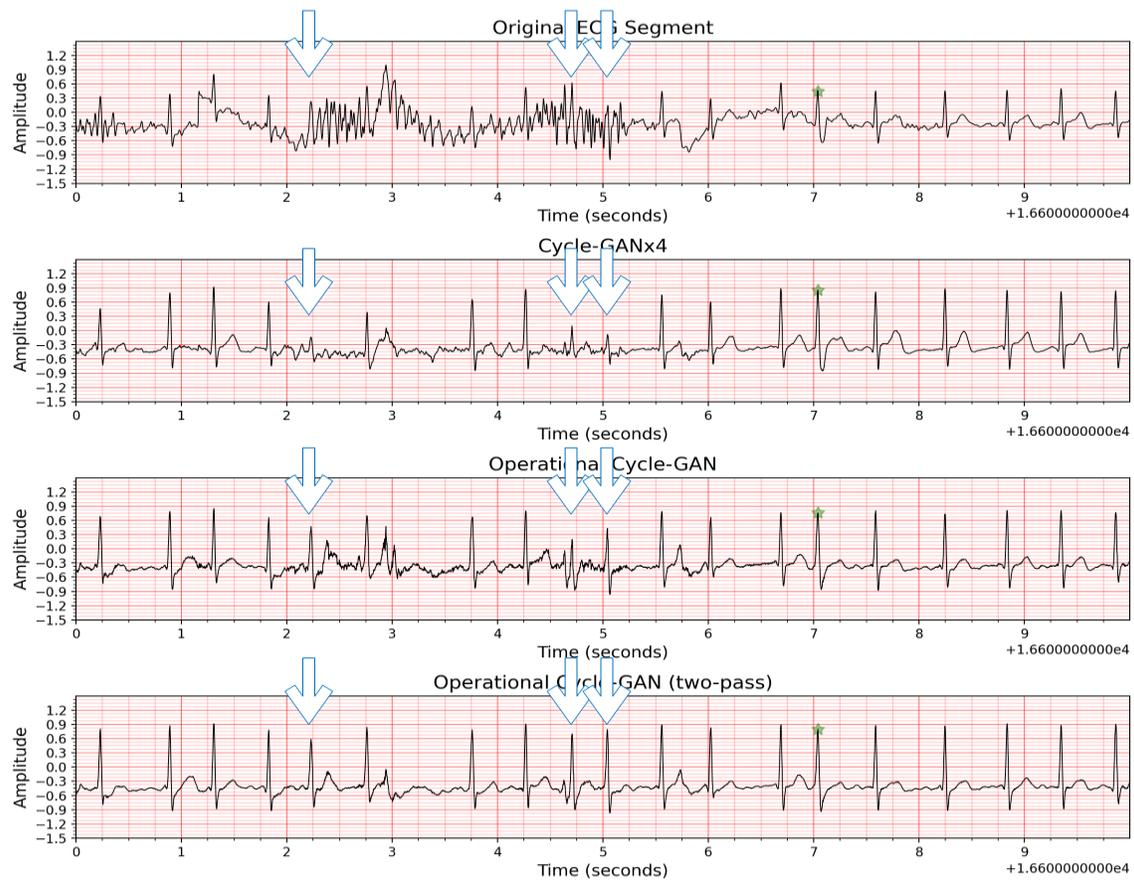

Figure 8: Sample ECG segment from Patient 2 and its corresponding GAN output signals.

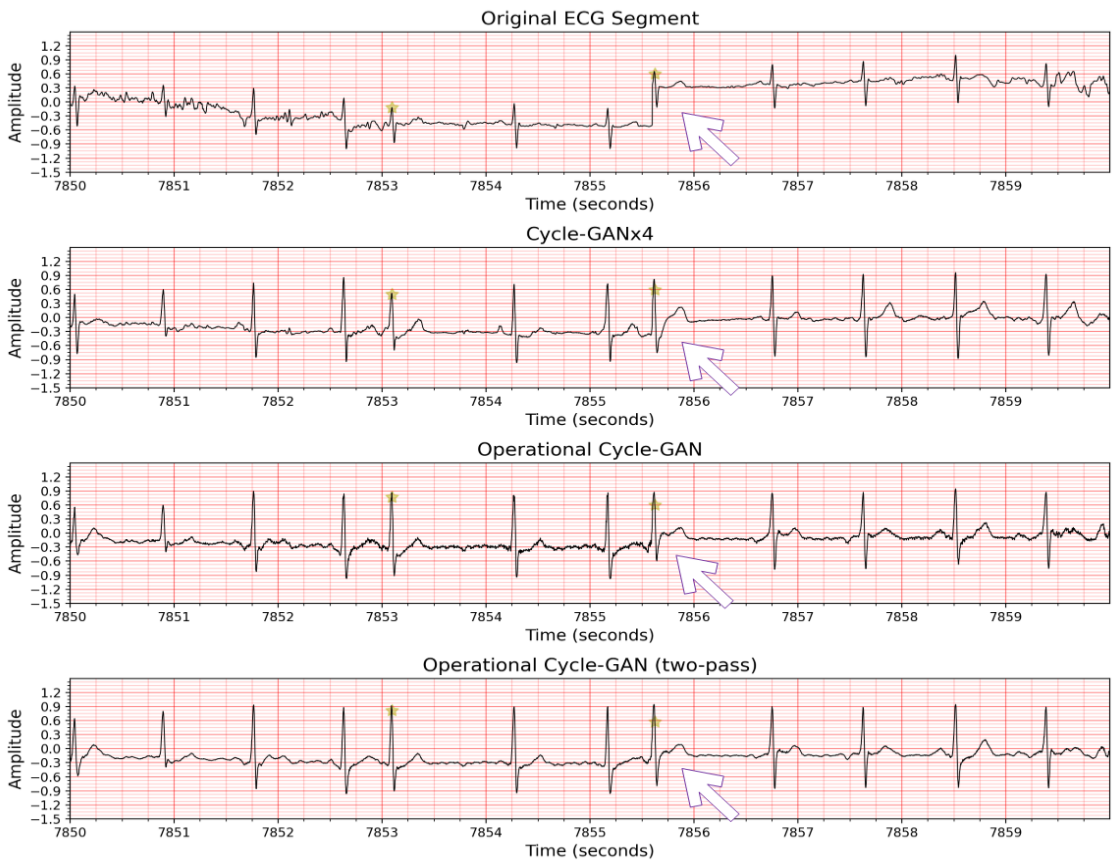

Figure 9: Sample ECG segment from Patient 10 and its corresponding GAN output signals.

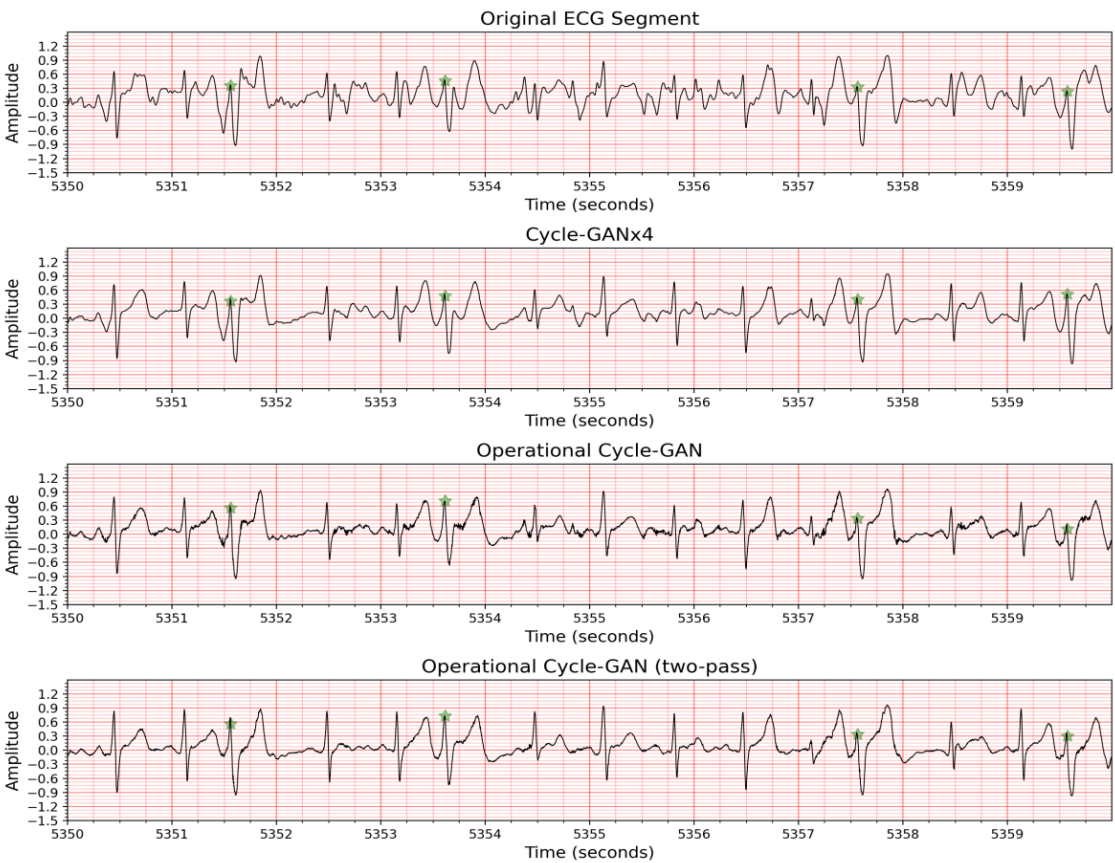

Figure 10: Sample ECG segment from Patient 7 and its corresponding GAN output signals.

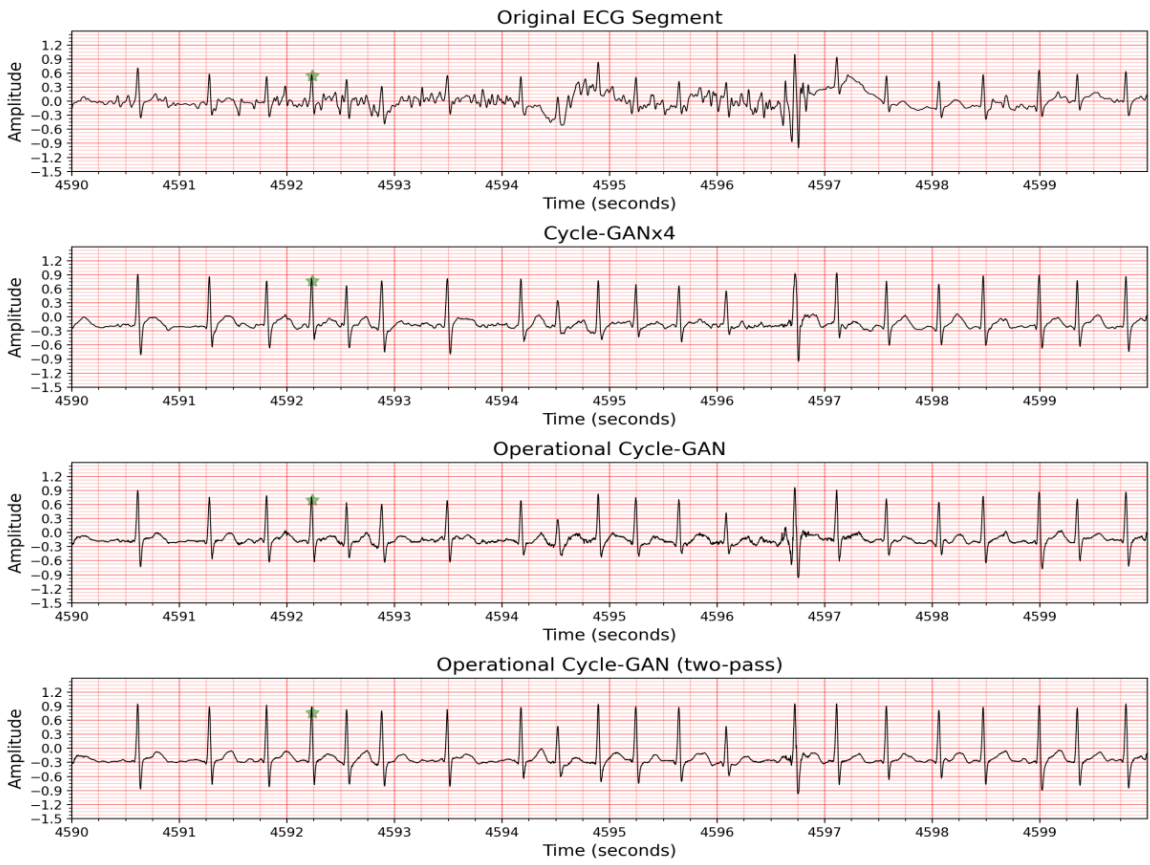

Figure 11: Sample ECG segment from Patient 2 and its corresponding GAN output signals.

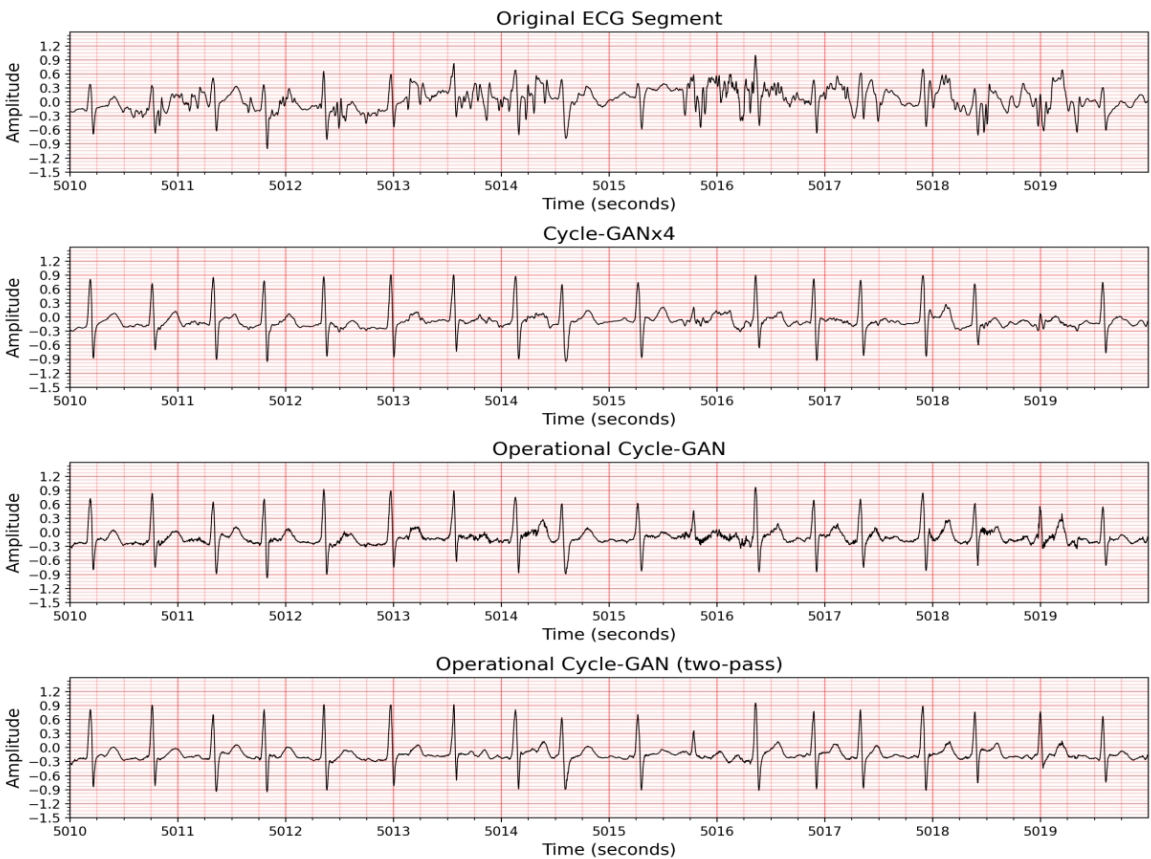

Figure 12: Sample ECG segment from Patient 3 and its corresponding GAN output signals.

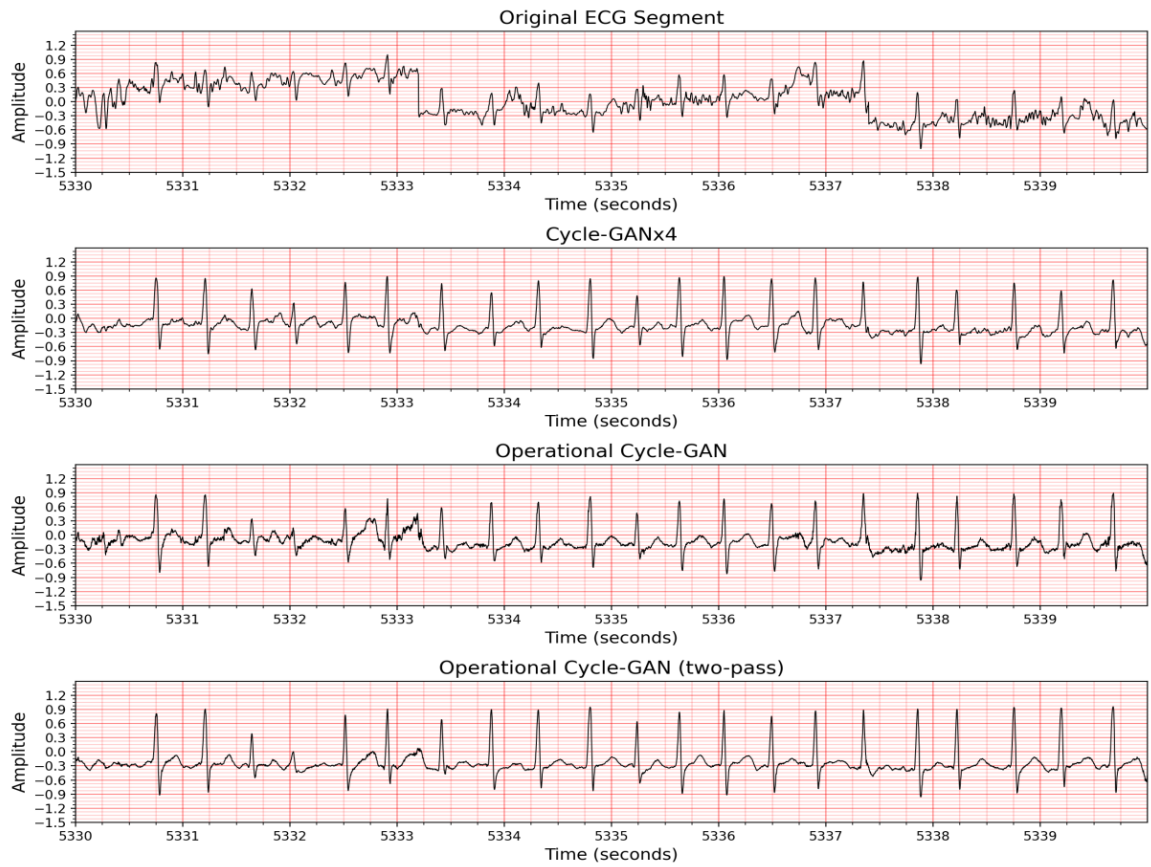

Figure 13: Sample ECG segment from Patient 3 and its corresponding GAN output signals.

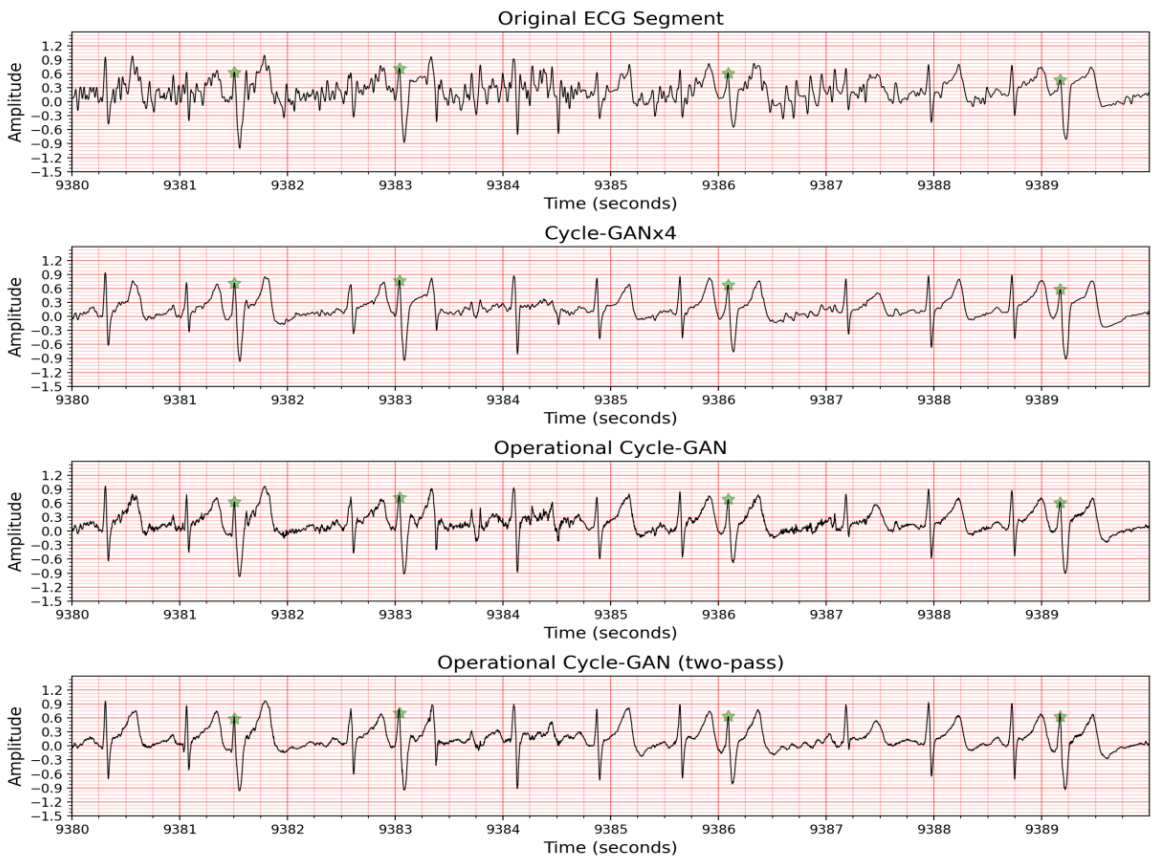

Figure 14: Sample ECG segment from Patient 7 and its corresponding GAN output signals.

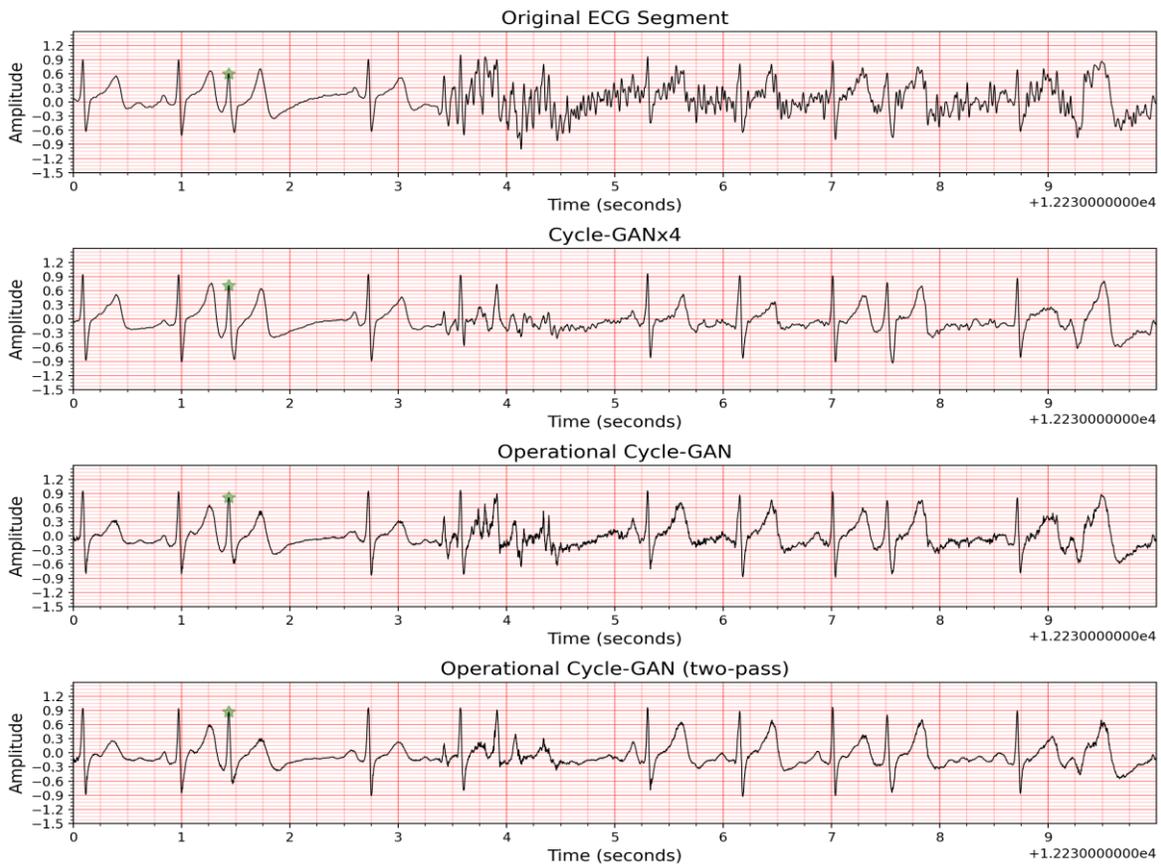

Figure 15: Sample ECG segment from Patient 7 and its corresponding GAN output signals.

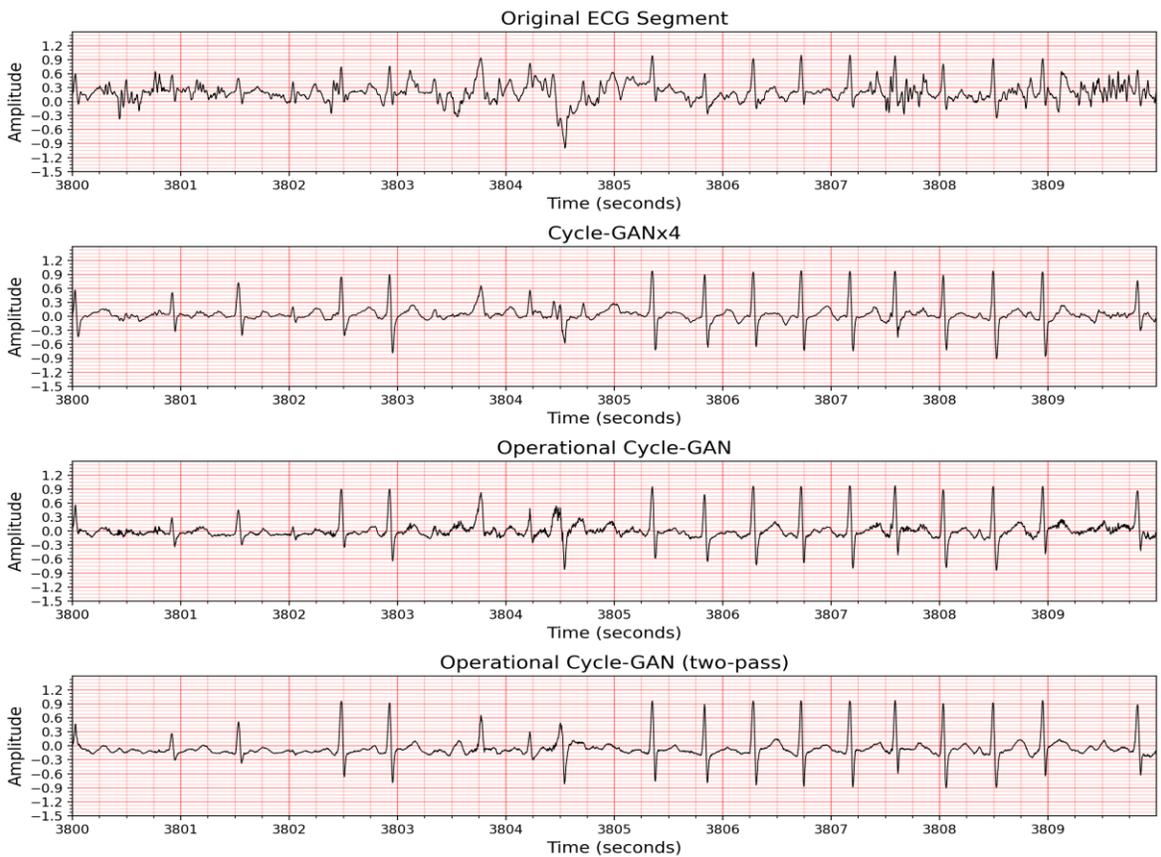

Figure 16: Sample ECG segment from Patient 8 and its corresponding GAN output signals.

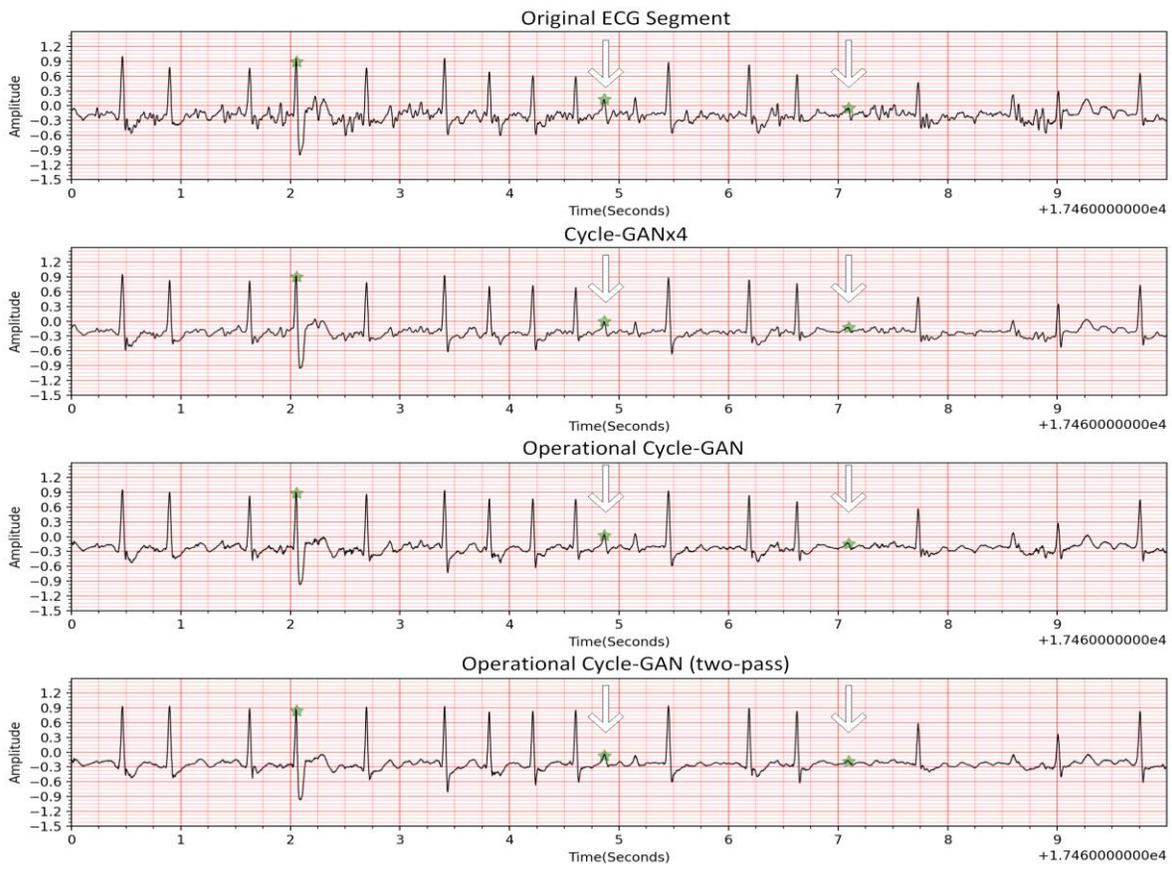

Figure 17: The arrows show occasional restoration failures when the arrhythmia peak is too low.

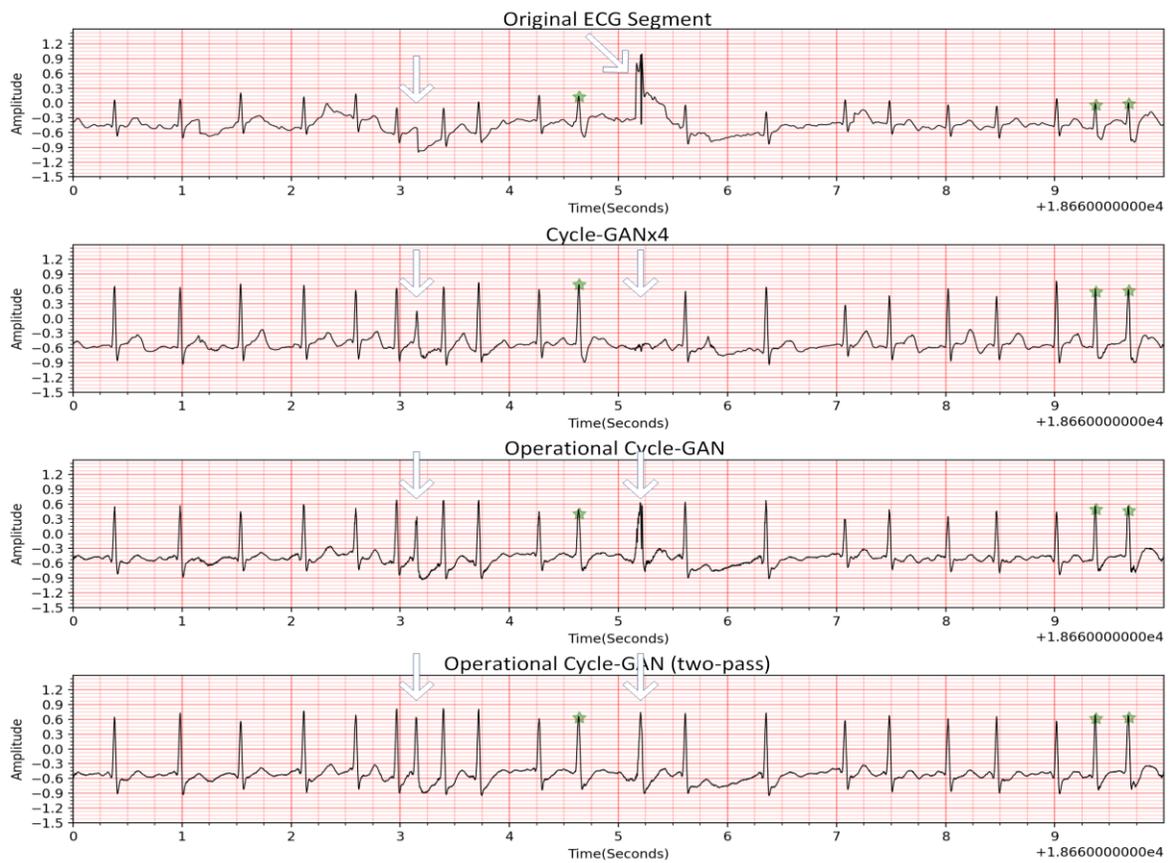

Figure 18: The arrows show occasional over-corrections.